# Speech & Song Emotion Recognition Using Multilayer Perceptron and Standard Vector Machine

Dr Behzad Javaheri

Department of Computer Science, City University of London, London, UK. behzad.javaheri@city.ac.uk

**Abstract**: herein, we have compared the performance of SVM and MLP in emotion recognition using speech and song channels of the RAVDESS dataset. We have undertaken a journey to extract various audio features, identify optimal scaling strategy and hyperparameter for our models. To increase sample size, we have performed audio data augmentation and addressed data imbalance using SMOTE. Our data indicate that optimised SVM outperforms MLP with an accuracy of 82 compared to 75%. Following data augmentation, the performance of both algorithms was identical at ~79%, however, overfitting was evident for the SVM. Our final exploration indicated that the performance of both SVM and MLP were similar in which both resulted in lower accuracy for the speech channel compared to the song channel. Our findings suggest that both SVM and MLP are powerful classifiers for emotion recognition in a vocal-dependent manner.

**Introduction and motivation**: emotion recognition is an important problem receiving increasing research interest due to its numerous applications in many domains, including audio surveillance, E-learning, clinical studies, lie detection and entertainment. This is a challenging task for machine learning primarily due to the uncertainty surrounding important predictor selection [1].

Emotion recognition attempts to decipher and classify a person's expression of mental state and perspective, evident from changes in one or more emotion modalities including facial expression, voice, body language as well as changes in brain signal captured by electroencephalography. Humans naturally build the capacity to identify emotions, though development, machines, however, find it difficult to access depth behind the content, therefore need to construct instantaneous emotion classification capability [2].

Recent advances in deep learning have allowed improvements in recognition problems such as face, voice, image and speech emotion recognition mainly through two approaches. In one approach, raw sound files are utilised as inputs hoping that neural networks detect significant predictors directly from the raw samples [3]. Alternatively, one or more extracted representation of a sound file used as input [4, 5]. These studies have used different predictors for emotion recognition and consensus for what predictors influence emotion recognition is lacking. Previous studies indicate that prosodic parameters including intensity, fundamental frequency and speaking rate are potentially strong emotion predictors for audio files. Voice quality and short-term spectral predictors have also been reported as effective predictors. There is also considerable uncertainty as to the best algorithm for emotion classification and whether one algorithm can sufficiently predict emotions from combined audio files of a different vocal channel; for example song and speech [6].

In this work, we take a holistic approach and use 5 different predictors from combined speech and song files using the RAVDESS dataset [7] and utilise Standard Vector machine (SVM) and Multilayer Perceptron (MLP) to build models with and without optimal hyperparameters. We suggest that increasing the number of predictors in input data may provide a better representation of a sound file, potentially leading to greater generalisation and higher performance of emotion recognition. In addition, we utilise data augmentation to increase sample size and examine whether classification performance may differ based on vocal channel.

**Description of the dataset**: the Ryerson Audio-Visual Database of Emotional Speech and Song (RAVDESS) was utilised in this study. The audio files recorded in speech or song vocal channels by 24 (12 males, 12 females) and 23 (12 males, 11 females) actors in each channel, respectively and are available in .wav format. The emotions expressed within the speech channel include calm, happy, sad, angry, fearful, disgust and surprise, and for the song include calm, happy, sad, angry, and fearful. The speech channel has 1440 (60 trials per actor x 24 actors) and song channel 1012 (44 trials per actor x 23 actors) [7]. To avoid data imbalance, disgust and surprising emotions were dropped as these were only available in the speech and absent in the song channel This resulted in 2068 audio files for our analysis.

The combined speech & song files used for predictor extraction using Librosa library to extract 5 predictors: Mel-scaled spectrogram, Mel-frequency Cepstral Coefficients (MFCCs), Chromagram, Spectral contrast and Tonnetz representation. The utility of several different audio predictors, instead of just one, allows a better representation of different sound characteristics. This leads to a richer description of a sound sample, which may improve the performance of speech and song related emotion recognition [8]. Each of these predictors can produce many numerical columns, for example, in our studies, MFCC produces 40 columns. A total of 194 columns related to the above predictors extracted.

**SVM and MLP:** SVM is a powerful deterministic supervised learning algorithm for both classification and regression with good generalisation and able to recognise patterns of non-linear relationships. It constructs an optimal hyperplane in which the margin between the classes is maximised [9]. It is a two-layer neural network employing a hidden layer of radial units and one output neuron. The construct of this network and its parameters is governed by kernel functions instead of direct processing of hidden unit signals [10]. SVM is considered a "shallow" architecture algorithm with few parameters to tune and therefore suffer from over-reliance on kernel function with very limited flexibility. Some of the advantages of SVM over MLP linked to the complexity of networks. SVMs use a large number of learning problem formulations leading to solving a quadratic optimisation problem [11, 12]. SVM can perform high by utilising structural risk minimisation rather than empirical risk minimisation inductive principle [13].

MLP is non-deterministic and universal in the sense that it can approximate any continuous nonlinear function sufficiently well on a compact interval and is used for various tasks including classification and regression [14, 15]. In MLP the neurons are arranged in layers, counting from the input layer to hidden layers ultimately to the output layer. The two neighbouring layers can connect and the network is feedforward [16]. Essentially, MLP contains several highly interconnected processing neurons that can be used in parallel to detect a solution for a given problem as an input. However, one of the MLP's drawbacks is that it may underperform when it tries to solve a nonlinear optimisation problem with many local minima. Furthermore, it may overfit on small datasets and as such sensitive to size of data [16, 17].

In this study, the robustness of MLP and SVM were evaluated for speech and song emotion recognition. Additionally, we examined whether their performances are dependent on vocal channels.

**Hypothesis**: that the SVM and MLP will produce highly accurate emotion classification in a vocal-independent manner.

**Description and choice of training and evaluation methodology**: data distribution was checked to identify whether scaling was required and find significant variations in data distribution. It was therefore decided to scale the data. Initially, the predictors (195) and the target variable (emotions) were defined. From the sklearn library, the train_test_split function was used to split predictors and target into 80 and 20% for train and test subsets, respectively. Subsequently, to identify the most appropriate scaling approach, two available scaling methods within the sklearn library were used. StandardScaler (standard) achieves scaling by producing predictors with mean zero and scaling data to unit variance, that is, variance and standard deviation of 1. Min-Max scaling will transform predictors into a range between [0, 1] or [-1, 1]. Furthermore, the target multiclass variable (emotions) was in categorical form, therefore encoded using LabelEncoder. The following process was followed: **1)** To identify optimal scaling method SVML and ML (non-optimised) were tested on non-scaled, standard and MinMax scaled. This step resulted in 7 models (3 for SVM and 4 for MLP with/without early stopping). **2)** Hyperparameter search to identify optimal parameters were performed and the performance of this optimal SVM and MLP models were evaluated on the entire training as well as testing datasets. This step resulted in 2 models. **3)** Data augmentation (to increase sample size and introduce noise) and SMOTE (Synthetic Minority Over-sampling Technique) to balance the dataset performed. Hyperparameter search performed and performance of optimal SVM and MLP were examined. This resulted in 2 models. **4)** The data split into "speech" and "song" channels, hyperparameter tuning performed and performance of optimal SVM and MLP models evaluated on both vocal channels. This resulted in 4 models.

To visualise and compare the performance of these models, cross-validation score, learning curve, classification report, class prediction error, precision-recall, ROC/AUC graph and confusion matrix were plotted using Yellowbrick library. Additionally, for both SVM and MLP search, the computing time was recorded and the models in the search were visualised using Matplotlib's Axes3D, by plotting a 3D graph.

**Choice of parameters**: for hyperparameter tuning exhaustive grid search was performed using sklearn's GridSearchCV, however, this was terminated as after 48 hours without completion. Alternatively, sklearn's RandomizedSearchCV was performed for kernel (rbf, poly and linear), C (uniform distribution 2, 50), gamma (uniform distribution 0.01, 1) and 3 StratifiedKfold CV. For MLP the grid included search for hidden_layer_size [(8,), (180,), (300,), (100,50,), (10,10,10)], activation ['tanh', 'relu', 'logistic'], solver ['sgd', 'adam'], alpha [0.0001, 0.001, 0.01], epsilon [1e-08, 0.1], and learning rate ['adaptive', 'constant']. The optimal parameters based on accuracy for both SVM and MLP (with/without early stopping) used for emotion recognition on entire training dataset and performance of these models plotted using 6 graphs in total as described above.

**Results, analysis and critical evaluation**: SVM (non-optimised) on non-scaled data produced an accuracy of 33.2 and 28.7% on training/testing datasets respectively, with complete misclassification for angry class, with an average 0.59 precision and 0.69 AUC. The accuracy of SVM with MinMax scaling increased to 67.5 and 67.3% for training and testing datasets, with an average precision of 0.74 and 0.90 AUC. The performance of SVM on Standard scaled data show an accuracy of 75.76 and 70.77% for training and testing datasets. The result for grid search suggests (computing time of 53.2s) to use 41.8 for C, 0.03 for gamma and rbf as the kernel. The performance of optimal SVM on entire training and testing datasets indicate an accuracy of 100 and 82.3%, respectively with a mean CV score of 0.742, average precision of 0.91 and AUC of 0.96. Our data indicate that neutral and calm classes had high correct classification evident from precision, recall, F and confusion matrix (Fig. 1A-C top panels). The class imbalance for angry emotion is evident.

We then used a similar approach for MLP. The performance of MLP on non-scaled data indicate accuracies of 65.1 and 53.3 for training and testing datasets. The average precision was 0.70 with AUC of 0.89. The performance was equally low across all emotions with the lowest in sad and happy emotions. MLP's performance on Min-Max scaled data result in 79.8 and 68.6% accuracy for training and testing with an average precision of 0.78 and AUC of 0.94. In addition, MLP's performance on the standard scaled data produced 100% accuracy for the training 75% for the testing dataset. The average precision was 0.87 and AUC was reported as 0.96. Visualisation of class prediction error suggests that happy and fearful emotions have the highest misclassifications. Grid search computed in 260.7 seconds suggested using "adam" as solver, a learning rate of "constant, hidden layer size of (100, 50) epsilon of 1e-08, alpha of 0.001 and "relu" as activation function. The optimal MLP leads to 100% accuracy for training and 76.33% for testing dataset with a mean cross-validation score of 0.737. Visualisation of the learning curve indicates overfitting. Precision was reported as 0.86 and AUC as 0.96 with happy emotion with the lowest and neutral emotion with the highest precision, recall and F1. To address overfitting MLP was re-run with earl stopping which resulted in accuracy of 87 and 74.64% for training and testing datasets with an average precision of 0.83 and 0.95 for AUC. Our data indicate that happy emotion has the lowest performance with 0.66 for precision, recall and F1 and having 24 out of 71 samples misclassified mostly within sad and fearful emotions. Neutral and calm classes enjoyed the higher true positive and lower true negatives (Fig. 1A-C bottom panels).

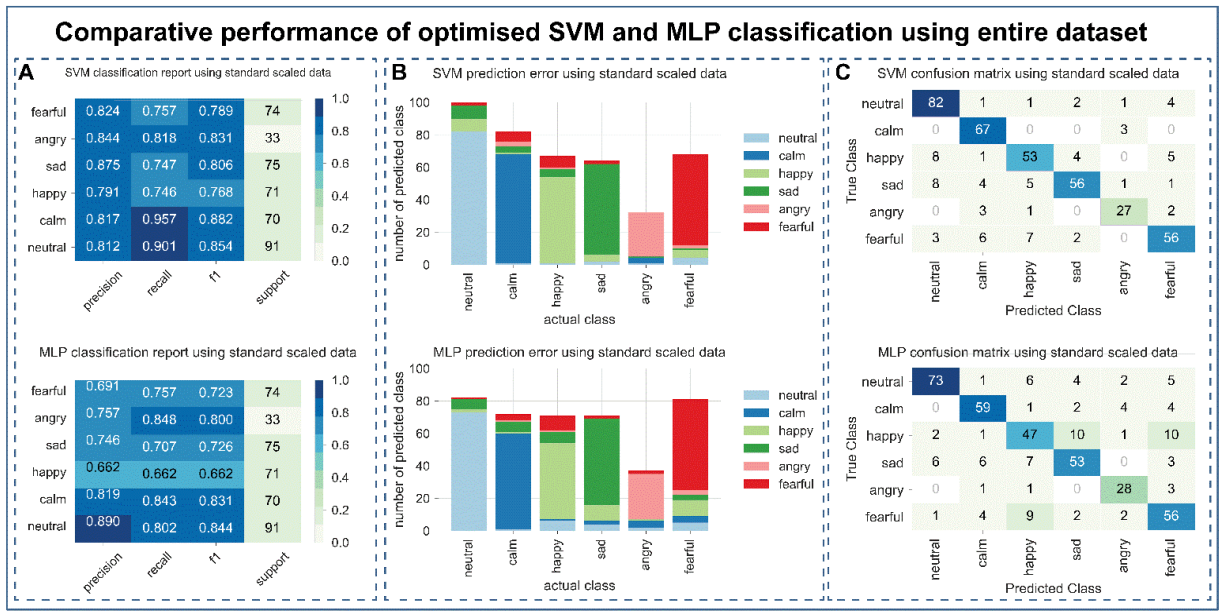

**Figure 1. Comparative performance of SVM and MLP emotion classification.** A) Classification report; B) prediction error and C) confusion matrix of optimised SVM & MLP emotion classification.

To increase sample number, introduce noise and overcome overfitting and class imbalance observed particularly for angry emotion, two approaches were employed. To increase the sample size and introduce noise, pydiogment library was employed to generate new audio files for each emotion based on the properties of audio files within each emotion. Two approaches of "fade in and out" and "change tone" were applied. To overcome class imbalance, SMOTE function of the imblearn library was utilised.

The computation time for the SVM hyperparameter search was 5203s with 44.9 for C, 0.155 gamma and poly as the kernel. The SVM performance was 100 and 79.1% on the entire training and testing datasets with a cross-validation score of 0.744, the precision of 0.78 and 0.93 for AUC (Figure 3). Examination of the learning curve indicates a straight line (score 1) across all training instances suggesting overfitting. Class prediction error show that data belonging to happy and fearful emotions were most contributory to misclassified emotions and neutral and angry were least contributory (Fig. 2 A-C top panels).

Similarly, hyperparameter tuning for MLP suggested utilising adam as the solver, adaptive learning rate and 300 hidden layer size, epsilon of 1e-8, alpha of 0.0001 and activation function of relu with a computation time of 703.2s. The performance of optimal MLP showed an accuracy of 93.5 and 78.8% for the training and testing dataset respectively. Examination of the learning curve did not suggest overfitting. Our data indicate an average cross-validation score of 0.725, average precision of 0.85 and AUC of 0.96. Classification reports suggest that neutral, angry and calm had the highest accurate emotion recognition with happy emotion being the least accurate with 0.67 precision, 0.680 recall and 0.675 f1 (Fig. 2 A-C bottom panel).

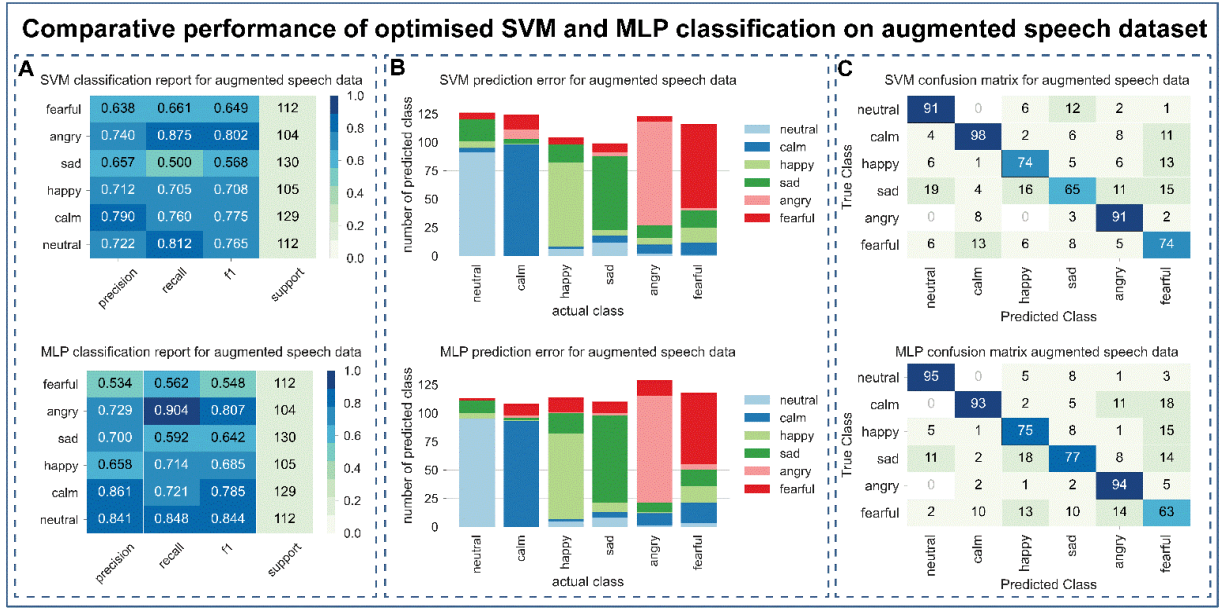

**Figure 2. Comparative performance of SVM and MLP emotion classification on augmented data.** A) Classification report; B) prediction error and C) confusion matrix of optimised SVM and.

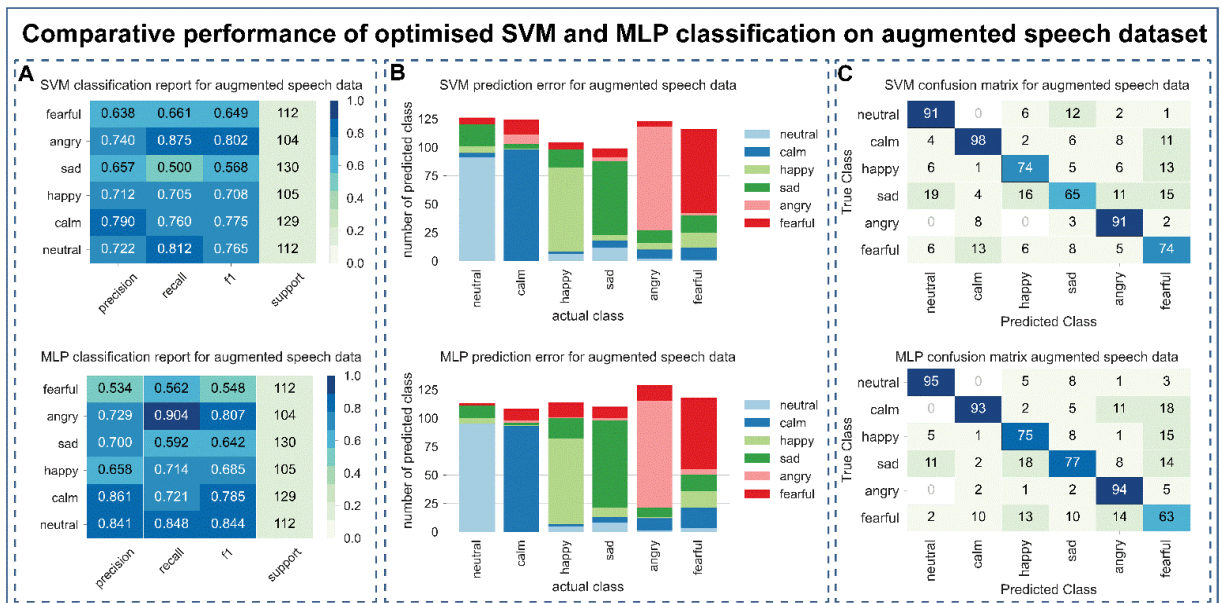

**Figure 3. Comparative performance of SVM and MLP classification on speech channel.** A) Classification report; B) prediction error & C) confusion matrix of optimised SVM and MLP classification using entire dataset.

To examine whether data split by vocal channel affects classification, data split into speech and song. Following hyperparameter tuning, optimised SVM model of the speech channel resulted in 100 and 71.24% accuracy for training/testing datasets with 0.632 average cross-validation score, average precision of 0.67 and AUC of 0.90. The learning curve suggests overfitting with a score 1 across all training instances. Classification report indicates sad emotion with 0.657 precision, 0.5 recall and 0.568 F1 had the lowest positive prediction (Fig. 3A-C top panel). For the song channel, accuracy of SVM on training/testing datasets were 100 and 88.9 respectively with learning curve score of 1 similarly indicating overfitting. The average CV score was 0.847, with an average precision of 0.92 and 0.96 for AUC. Angry emotion scored high with 1 for precision, 0.991 for recall and 0.995 for F1. The least accurate was for happy with 0.778, 0.840 and 0.808 for precision, recall and F1 (Fig. 4A-C). Furthermore, emotion recognition in speech channel using optimised MLP resulted in 95 and 71.8% accuracy for train/test datasets. The average CV score was 0.620 with average precision of 0.67 and 0.94 AUC. Classification report & class prediction error indicate that neutral emotion had the highest true positive & lowest false positive with a precision of 0.841, recall of 0.848 and 0.844 F1. Angry emotion performed similarly well. In contrast, fearful emotion had the lowest true positives & highest false positives with 0.534 precision, 0.562 recall and 0.548 F1 (Fig. 3A-C). In contrast, the performance of optimised MLP on the song channel was higher with 96.45 and 85.07% accuracy for train/test datasets with 0.804 average CV score, average precision of 0.88 and 0.98 AUC. The angry emotion had the highest true positive and lowest false positives with 0.991 precision, 0.972 recall and 0.981 F1 (Fig. 4A-C bottom panels).

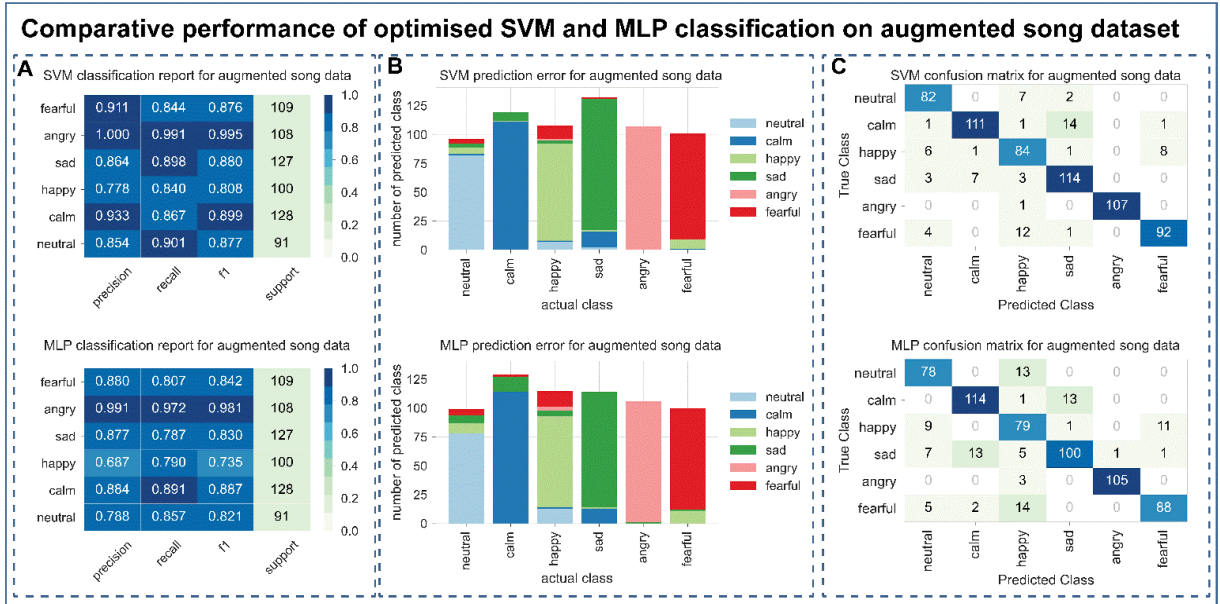

**Figure 4. Comparative performance of SVM and MLP # classification on song channel.** A) Classification report; B) prediction error & C) confusion matrix of optimised SVM and MLP classification using entire dataset.

**Conclusions**: In the present work we have performed a comparative study on the performance of SVM and MLP to classify 6 emotions from audio file datasets. We have taken a holistic approach by extracting 5 predictors (many inputs from each) from the audio files and have undertaken a journey whereby we compared the performance of a) SVM and MLP on non-scaled, MinMax and standard scaled dataset; b) optimised SVM and MLP on a standard scaled dataset; c) optimised SVM and MLP on augmented and balanced (SMOTE) dataset and d) optimised SVM and MLP on each speech and song channels. This resulted in 8 grid searches to find optimal hyperparameters and 15 models in total. Our data show that hyperparameter tuning increased SVM's performance and that augmentation and addressing data imbalance does not seem to further increase this performance. Optimisation addressed MLP's overfitting but did not further enhance accuracy for the test subset and that data augmentation and addressing imbalance enhances MLP's accuracy. Interestingly, when data split by vocal channel, both SVM and MLP performed similarly for both speech and song channels but SVM exhibited overfitting as evident by learning curve whereas this was not observed for MLP. Additionally, both SVM and MLP performed better in emotion classification using the song compared to speech channel.

Previous studies on emotion recognition from audio files have used a variety of datasets and approaches. For example, Gideon et al., (2017) performed a comparative study of the performance of progressive neural networks to conventional deep neural network using the IEMOCAP and MSP-IMPROV datasets aiming to devise a multi-task learning approach for emotion, gender and speaker recognition [18]. In addition, Buyukyilmaz and Cibikdiken (2016) used the MLP model on a dataset (3,168 files of male and female voices) for gender classification and reported 96.74% accuracy [19]. Furthermore, Shegokar and Sircar (2016) utilised SVM for the classification of male speech samples from the RAVDESS dataset and obtained an accuracy of 60.1% [20]. Zhao *et al.*, (2019) report 95.8% accuracy using a subsection of the dataset and convolutional neural network/long short-term memory [21]. More relevant to this work, Popova *et al.*, (2017) reported 71% accuracy in emotion recognition using a convolutional neural network on the RAVDESS dataset [22]. Similar to this work, Joy *et al.*, (2020) compared SVM and MLP's performance in emotion recognition and reported accuracy of 70 and 58%, respectively [23]. Our best-performing model reports higher accuracy in emotion recognition with the directly comparable studies. It also outperforms a recent study in which 71.6% accuracy in emotion recognition using the RAVDESS dataset was reported [8].

**Lessons learned:** both SVM and MLP are powerful classifiers for emotion recognition with vocal-dependent performance in which better classification is observed in the song compared to speech channel.

**Future work:** The task of emotion recognition from voice predictors differ between males and females [24, 25], therefore the effect of gender on classification performance can be further explored. In addition, ageing also affects the acoustical voice characteristics [26] and thus this important contributor has to be considered for refinement of classification strategies.

**References**

[1] A. M. Badshah, J. Ahmad, N. Rahim, and S. W. Baik, "Speech emotion recognition from spectrograms with deep convolutional neural network," in *2017 international conference on platform technology and service (PlatCon)*, 2017: IEEE, pp. 1-5.

[2] S. Brave and C. Nass, "Emotion in human–computer interaction," in *Human-computer interaction fundamentals*, vol. 20094635: CRC Press Boca Raton, FL, USA, 2009, pp. 53-68.

[3] G. Trigeorgis *et al.*, "Adieu features? end-to-end speech emotion recognition using a deep convolutional recurrent network," in *2016 IEEE international conference on acoustics, speech and signal processing (ICASSP)*, 2016: IEEE, pp. 5200-5204.

[4] K. Han, D. Yu, and I. Tashev, "Speech emotion recognition using deep neural network and extreme learning machine," in *Fifteenth annual conference of the international speech communication association*, 2014.

[5] W. Lim, D. Jang, and T. Lee, "Speech emotion recognition using convolutional and recurrent neural networks," in *2016 Asia-Pacific signal and information processing association annual summit and conference (APSIPA)*, 2016: IEEE, pp. 1-4.

[6] M. Pantic and L. J. Rothkrantz, "Toward an affect-sensitive multimodal human-computer interaction," *Proceedings of the IEEE,* vol. 91, no. 9, pp. 1370-1390, 2003.

[7] S. R. Livingstone and F. A. Russo, "The Ryerson Audio-Visual Database of Emotional Speech and Song (RAVDESS): A dynamic, multimodal set of facial and vocal expressions in North American English," *PloS one,* vol. 13, no. 5, p. e0196391, 2018.

[8] D. Issa, M. F. Demirci, and A. Yazici, "Speech emotion recognition with deep convolutional neural networks," *Biomedical Signal Processing and Control,* vol. 59, p. 101894, 2020.

[9] L. Auria and R. A. Moro, "Support vector machines (SVM) as a technique for solvency analysis," 2008.

[10] S. Osowski, K. Siwek, and T. Markiewicz, "Mlp and svm networks-a comparative study," in *Proceedings of the 6th Nordic Signal Processing Symposium, 2004. NORSIG 2004.*, 2004: IEEE, pp. 37-40.

[11] L. Feng *et al.*, "Alfalfa Yield Prediction Using UAV-Based Hyperspectral Imagery and Ensemble Learning," *Remote Sensing,* vol. 12, no. 12, p. 2028, 2020.

[12] M. Hesami, R. Naderi, M. Tohidfar, and M. Yoosefzadeh-Najafabadi, "Development of support vector machine-based model and comparative analysis with artificial neural network for modeling the plant tissue culture procedures: effect of plant growth regulators on somatic embryogenesis of chrysanthemum, as a case study," *Plant Methods,* vol. 16, no. 1, pp. 1-15, 2020.

[13] A. Belayneh, J. Adamowski, B. Khalil, and B. Ozga-Zielinski, "Long-term SPI drought forecasting in the Awash River Basin in Ethiopia using wavelet neural network and wavelet support vector regression models," *Journal of Hydrology,* vol. 508, pp. 418-429, 2014.

[14] S. K. Pal and S. Mitra, "Multilayer perceptron, fuzzy sets, classifiaction," 1992.

[15] D. E. Rumelhart, G. E. Hinton, and R. J. Williams, "Learning representations by back-propagating errors," *nature,* vol. 323, no. 6088, pp. 533-536, 1986.

[16] M. A. Ghorbani, H. A. Zadeh, M. Isazadeh, and O. Terzi, "A comparative study of artificial neural network (MLP, RBF) and support vector machine models for river flow prediction," *Environmental Earth Sciences,* vol. 75, no. 6, p. 476, 2016.

[17] M. Geetha, "Forecasting the Crop Yield Production in Trichy District Using Fuzzy C-Means Algorithm and Multilayer Perceptron (MLP)," *International Journal of Knowledge and Systems Science (IJKSS),* vol. 11, no. 3, pp. 83-98, 2020.

[18] J. Gideon, S. Khorram, Z. Aldeneh, D. Dimitriadis, and E. M. Provost, "Progressive neural networks for transfer learning in emotion recognition," *arXiv preprint arXiv:1706.03256,* 2017.

[19] M. Buyukyilmaz and A. O. Cibikdiken, "Voice gender recognition using deep learning," in *2016 International Conference on Modeling, Simulation and Optimization Technologies and Applications (MSOTA2016)*, 2016: Atlantis Press, pp. 409-411.

[20] P. Shegokar and P. Sircar, "Continuous wavelet transform based speech emotion recognition," in *2016 10th International Conference on Signal Processing and Communication Systems (ICSPCS)*, 2016: IEEE, pp. 1-8.

[21] J. Zhao, X. Mao, and L. Chen, "Speech emotion recognition using deep 1D & 2D CNN LSTM networks," *Biomedical Signal Processing and Control,* vol. 47, pp. 312-323, 2019.

[22] A. S. Popova, A. G. Rassadin, and A. A. Ponomarenko, "Emotion recognition in sound," in *International Conference on Neuroinformatics*, 2017: Springer, pp. 117-124.

[23] J. K. Joy, Aparna; Ram, Shreya; Rama, S, "Speech Emotion Recognition using Neural Network and MLP Classifier," *IJESC,* vol. 10, no. 4, 2020.
[24] D. H. Klatt and L. C. Klatt, "Analysis, synthesis, and perception of voice quality variations among female and male talkers," *the Journal of the Acoustical Society of America,* vol. 87, no. 2, pp. 820-857, 1990.
[25] I. R. Titze, "Physiologic and acoustic differences between male and female voices," *The Journal of the Acoustical Society of America,* vol. 85, no. 4, pp. 1699-1707, 1989.
[26] J. Sundberg, M. N. Thörnvik, and A. M. Söderström, "Age and voice quality in professional singers," *Logopedics Phoniatrics Vocology,* vol. 23, no. 4, pp. 169-176, 1998.